\documentclass[pdflatex,iicol,sn-mathphys-num]{sn-jnl}

\usepackage{graphicx}%
\usepackage{multirow}%
\usepackage{amsmath,amssymb,amsfonts}%
\usepackage{amsthm}%
\usepackage{mathrsfs}%
\usepackage[title]{appendix}%
\usepackage{xcolor}%
\usepackage{textcomp}%
\usepackage{manyfoot}%
\usepackage{booktabs}%
\usepackage{algorithm}%
\usepackage{algorithmicx}%
\usepackage{algpseudocode}%
\usepackage{listings}%

\usepackage{xspace}
\usepackage{bm}
\usepackage[capitalise]{cleveref}
\usepackage{maybemath}
\usepackage{caption}
\usepackage{subcaption}

\makeatletter 
\def\@acknow{}%
\long\def\EarlyAcknow#1 \par{%
\def\@acknow{\abstractfont\abstracthead*{Acknowledgments}
#1\par}}%

\def\printabstract{\ifx\@acknow\empty\else\@acknow\fi\par%
    \ifx\@abstract\empty\else\@abstract\fi\par}
\makeatother

\DeclareRobustCommand{\thesismath}[1]{\ensuremath{\maybebmsf{#1}}}
\newcommand{\rcite}[1]{Ref.~\cite{#1}\xspace}

\newcommand*{\bcjets}{\ensuremath{b}- and \ensuremath{c}-jets\xspace}
\newcommand*{\bhadron}{\ensuremath{b\text{-hadron}}\xspace}
\newcommand*{\bhadrons}{\ensuremath{b\text{-hadrons}}\xspace}
\newcommand*{\chadron}{\ensuremath{c\text{-hadron}}\xspace}
\newcommand*{\chadrons}{\ensuremath{c\text{-hadrons}}\xspace}
\newcommand*{\bjet}{\ensuremath{b\text{-jet}}\xspace}
\newcommand*{\bjets}{\ensuremath{b\text{-jets}}\xspace}

\newcommand*{\cjets}{\ensuremath{c\text{-jets}}\xspace}

\newcommand*{\ljets}{light-jets\xspace}
\DeclareRobustCommand{\dr}{\thesismath{dR}\xspace}
\newcommand*{\crej}{\ensuremath{c}-jet rejection\xspace}
\newcommand*{\lrej}{light-jet rejection\xspace}
\DeclareRobustCommand{\lxy}{\thesismath{L_{xy}}\xspace}
\newcommand{\maskformer}{MaskFormer\xspace}

\newcommand\pct[1]{\ensuremath{#1\%}\xspace}
\DeclareRobustCommand{\pt}{\thesismath{p_{\mathrm{T}}}\xspace}
\newcommand{\NULL}{\textsc{null}\xspace}

\begin{document}

\title[Secondary Vertexing with MaskFormers]{Secondary Vertex Reconstruction with MaskFormers}


\author*[1]{\fnm{Samuel} \sur{Van Stroud}
} \email{sam.van.stroud@cern.ch}
\author[1]{\fnm{Nikita} \sur{Pond} 
}\email{nikita.pond.18@ucl.ac.uk}
\author[1]{\fnm{Max} \sur{Hart} }\email{max.hart.22@ucl.ac.uk}
\author[1,2]{\fnm{Jackson} \sur{Barr} 
}\email{jackson.barr.17@ucl.ac.uk}
\author[1,3]{\fnm{S\'ebastien} \sur{Rettie} 
}\email{s.rettie@ucl.ac.uk}
\author*[1]{\fnm{Gabriel} \sur{Facini} 
}\email{g.facini@ucl.ac.uk}
\author[1]{\fnm{Timothy} \sur{Scanlon} 
}\email{timothy.scanlon@ucl.ac.uk}

\affil[1]{\orgdiv{Centre for Data Intensive Science and Industry}, \orgname{University College London}}

\affil[2]{\orgname{Deutsches Elektronen-Synchrotron DESY}}

\affil[3]{\orgname{CERN}}


\EarlyAcknow{We gratefully acknowledge the support of the UK's Science and Technology Facilities Council (STFC). N.P., M.H., and J.B. are supported under ST/W00674X/1 along with industry and departmental contributions. G.F. was supported by an Ernest Rutherford Fellowship (STN0039343), ST/W00058X/1, and STX0014071. T.S. receives support from the Royal Society (URF/R/180008) and STFC (ST/W00058X/1). S.R. is supported by CERN, the Banting Postdoctoral Fellowship program, and the Natural Sciences and Engineering Research Council of Canada. We also extend our thanks to UCL for the use of their high-performance computing facilities, with special thanks to Edward Edmondson for his expert management and technical support.}


\abstract{
In high-energy particle collisions, the reconstruction of secondary vertices from heavy-flavour hadron decays is crucial for identifying and studying jets initiated by \textit{b}- or \textit{c}-quarks.
Traditional methods, while effective, require extensive manual optimisation and struggle to perform consistently across wide regions of phase space.
Meanwhile, recent advancements in machine learning have improved performance but are unable to fully reconstruct multiple vertices.
In this work we propose a novel approach to secondary vertex reconstruction based on recent advancements in object detection and computer vision.
Our method directly predicts the presence and properties of an arbitrary number of vertices in a single model.
This approach overcomes the limitations of existing techniques.
Applied to simulated proton-proton collision events, our approach demonstrates significant improvements in vertex finding efficiency, achieving a 10\% improvement over an existing state-of-the-art method.
Moreover, it enables vertex fitting, providing accurate estimates of key vertex properties such as transverse momentum, radial flight distance, and angular displacement from the jet axis.
When integrated into a flavour tagging pipeline, our method yields a 50\% improvement in light-jet rejection and a 15\% improvement in \textit{c}-jet rejection at a \textit{b}-jet selection efficiency of 70\%.
These results demonstrate the potential of adapting advanced object detection techniques for particle physics, and pave the way for more powerful and flexible reconstruction tools in high-energy physics experiments.
}%
\keywords{Particle Physics, Vertex Reconstruction, Machine Learning, Object Detection, Flavour Tagging}
\maketitle%
\section{Introduction}\label{sec:intro}

Particle physics experiments at facilities such as the Large Hadron Collider (LHC) at CERN~\cite{LHC} have deepened our understanding of fundamental physics through the study of high-energy proton-proton ($pp$) collisions.
These collisions, observed in detectors like ATLAS~\cite{ATLAS-ATLAS} and CMS~\cite{CMS-CMS}, often result in the creation of resonances that decay into conical sprays of particles known as jets.
A critical component in analysing collisions is the ability to identify the origin of these jets, particularly those initiated by heavy-flavour quarks ($b$- or $c$-quarks)~\cite{ATLAS-FTag,CMS-FTag}.
This process, known as \textit{flavour tagging}, plays a crucial role in many physics analyses, from precision measurements of the Standard Model to searches for new phenomena.
Flavour tagging relies heavily on the reconstruction of charged particle trajectories (tracks).
In experiments like ATLAS and CMS, decays from \bhadrons typically produce four to five charged particles which must be distinguished from the roughly $10$--$100$ total particles within jets at the energies typically probed at the LHC.

Central to flavour tagging is the concept of a \textit{vertex}, which is a point where particles interact or decay.
The primary $pp$ collision location is reconstructed via a \textit{primary vertex}.
The decay points of heavy-flavour hadrons are identified as \textit{secondary vertices}.
Weakly decaying heavy-flavour hadrons have non-negligible lifetimes~\cite{PDG}, resulting in secondary decays that may be displaced from the initial $pp$ collision.
Within the general purpose detectors at the LHC, this displacement varies significantly depending on the hadron's lifetime and transverse momentum.
For the samples probed in this study, these displacements typically range from a negligible distance from the primary vertex to several centimetres into the detector.
Heavy-flavour jets typically contain one to three secondary vertices, while \ljets (those originating from light-flavour hadrons or gluons), typically contain secondary vertices at a lower rate and with a lower mass.
As a consequence, the reconstruction of secondary vertices within jets is a key component of flavour tagging.

A \textit{reconstructed vertex} refers to the point in space where particle tracks converge, indicating a possible interaction or decay point. 
Notably, a secondary vertex can be reconstructed with a single track as the jet axis often aligns with the heavy-flavour hadron trajectory.
Traditionally, vertex reconstruction occurs in two stages: vertex finding, which identifies the set of tracks with a compatible spatial origin, and vertex fitting, which estimates properties of the vertex from the set of compatible tracks.
However, existing procedural approaches for secondary vertex reconstruction often require extensive manual optimisation and may not perform consistently across wide regions of phase space.
Machine learning (ML) has been used in particle physics for several decades~\cite{firstmlftag}.
In recent years, the application of deep learning has become widespread and has led to dramatic performance improvements in various settings.
Existing state-of-the-art flavour tagging algorithms now fully embrace these methods and offer exceptional performance.
However, these methods either ignore vertex reconstruction altogether~\cite{ParticleNet, ParT}, or perform only a limited form of vertex finding without vertex fitting~\cite{GN1, GN2X}.

The main contribution of this article is to enable the full reconstruction (both finding and fitting) of multiple vertices inside a jet using a learned approach.
To achieve this, we propose a novel approach based on MaskFormer~\cite{maskformer}, which was originally designed for image segmentation and is based on the scalable Transformer architecture~\cite{VaswaniAttention}.
Our implementation processes unordered sets of input tracks and outputs a variable number of fully characterised vertices, each with a predicted class label, fitted properties, and track assignments.
By integrating this model with a state-of-the-art ML-based flavour tagging algorithm similar to one developed by ATLAS~\cite{GN1}, we demonstrate that the capacity to fully reconstruct heavy-flavour vertices yields a significant improvement in the ability to identify jets compatible with the decay of heavy-flavour hadrons.

The article is structured as follows:
Section~\ref{sec:related-work} reviews related work in high-energy physics and computer vision.
Section~\ref{sec:model} details our model architecture.
Section~\ref{sec:experiments} describes the simulated samples and training specifics.
Section~\ref{sec:results} presents results for vertex finding, fitting, and flavour tagging.
Section~\ref{sec:conclusion} summarises findings and discusses future applications.


\section{Related Work}
\label{sec:related-work}

Previous work in both high-energy physics and computer vision lays a robust foundation for our contributions.
We draw from the latest advancements in object detection and instance segmentation in computer vision to address the prevailing challenges in vertex reconstruction in high-energy physics, paving the way for more powerful physics analyses.

\subsection{High Energy Physics}\label{sec:related-hep}

High-energy physics experiments such as ATLAS and CMS rely heavily on primary and secondary vertex reconstruction algorithms for analysing collision events~\cite{Rudolf21}.
Traditional secondary vertex reconstruction algorithms~\cite{BTagPerf, JetFitter, CMS-BTV} work by iteratively clustering tracks around common points of origin and then extracting the vertex properties in a fit. 
While effective, these approaches require extensive manual optimisation and may not perform optimally across wide regions of phase space, such as the range of jet transverse momenta (\pt) produced in modern collider experiments.
This manual optimisation is complex, time-consuming, and challenging when deploying these algorithms in new environments.

The integration of ML techniques marks a significant shift in this domain. 
Existing ML-based secondary vertexing approaches use edge classification on input graphs to predict whether each pair of tracks in the jet belong to a common vertex~\cite{Shlomi:2020ufi,GN1}.
In this approach, compatible track pairs are used to form vertices by extracting the connected components of the resulting graph in a post-processing step.
As the formation of multi-track vertices is not directly learned, this approach may be suboptimal.
Additionally, edge classification cannot easily be used to estimate the properties (e.g. position and momentum) of an unknown number of vertices.
This technique is used in the EdgeClassifier benchmark model in \cref{sec:results}.

Other recent work includes differentiable fitting of a single inclusive secondary vertex~\cite{Smith:2023ssh}, and the application of CNNs for primary vertex identification~\cite{Fang:2019wsd,ATL-PHYS-PUB-2023-011,Akar:2023puf}.
The former approach is limited to the reconstruction of a single inclusive vertex per jet, while the latter does not provide track assignments and requires an expensive kernel density estimation step.
Our work overcomes these limitations by predicting the presence, properties, and track assignments of an arbitrary number of vertices in a single step.

\subsection{Computer Vision}

In computer vision, object detection is the task of identifying objects within an image and determining their properties, for example the type of object and its location within the image.
Object detection combines two tasks: semantic segmentation and instance segmentation. Semantic segmentation assigns a class label to each pixel, while instance segmentation differentiates between individual instances within the same semantic class.
Techniques in this field have evolved to predict object classes alongside either bounding boxes \cite{2015arXiv150602640R,2015arXiv150601497R,2020arXiv200512872C} and/or pixel masks \cite{2017arXiv170306870H,maskformer}.
The use of bounding boxes relies on the dense grid structure of images and the implicit ordering of their pixels.
However, methods relying on per-pixel mask prediction can readily be extended for sparse, unordered collections of inputs such as those found in high energy physics.

The \textit{Segment Anything} model \cite{segmentanything} is a state-of-the-art object detector based on a \maskformer architecture \cite{maskformer,mask2former}.
This architecture takes in an image and outputs a series of detected objects.
Each output object consists of a class label and a binary pixel mask.
The object classes are semantic labels for objects in images (e.g. ``dog'' or ``cat''), while the pixel masks are used to associate a set of pixels in the image to each labelled object.
Multiple instances of the same class are resolved as different masks with common class labels.
This concept of resolving multiple object instances of the same class is particularly relevant to our work.

Our research draws inspiration from these advancements in computer vision. 
We analogise a grid of image pixels to an unordered set of tracks within a jet, and image-depicted objects to heavy-flavour hadron decay vertices. 
This correspondence allows us to adapt computer vision techniques for particle physics, enabling predictions of both the presence and properties of multiple vertices in a novel and effective manner.


\section{Model Architecture}
\label{sec:model}
\begin{figure*}[t]
    \centering
    \includegraphics[width=\textwidth]{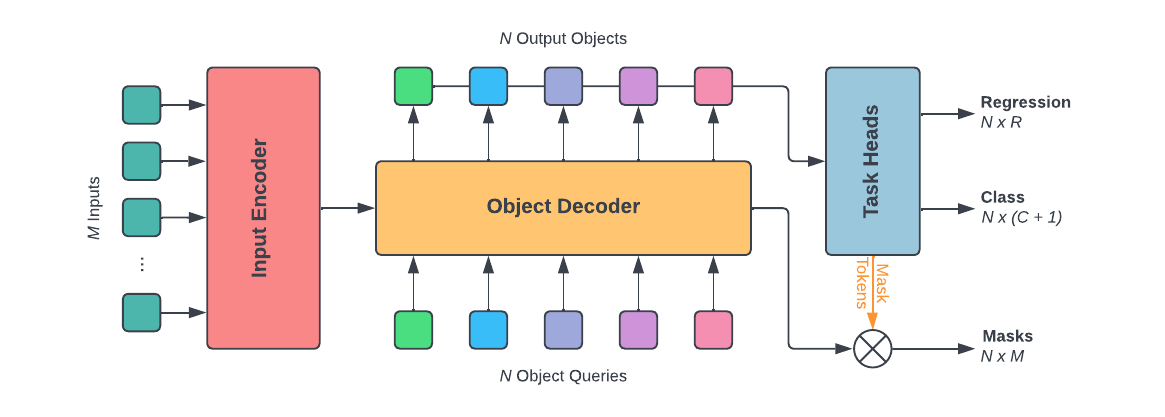}
    \caption{
        Overview of the \maskformer network architecture.
        $M$ input elements are fed into an initial Transformer encoder.
        The object decoder then takes a set of $N$ object queries and iteratively updates them with information from the input elements and other object queries.
        Finally, three task heads are used to: categorise each object as being from one of $C+1$ object classes (including a \NULL class), $R$ regression targets, and $N \times M$ binary masks which provide the assignment of input tracks to output vertices.
        The model can also be configured to predict a global class label, which is not shown here for simplicity.
    }
    \label{fig:high-level-arch}
\end{figure*}

Our approach is based on the \maskformer architecture as implemented in Refs. \cite{maskformer,mask2former,segmentanything} with the following modifications.
Firstly, the model is adapted to accept sets of charged particle tracks instead of images as inputs, resulting in a simpler Transformer backbone and the removal of upsampling in the mask generation stage.
Secondly, regression tasks are added to reconstruct the vertex properties.
Finally, to enable flavour tagging, the model is configured to output a prediction of the jet flavour.
The architecture is depicted in \cref{fig:high-level-arch} and described in more detail below.

\subsection{Input Encoder}
The first stage of the model is a dense embedding layer that outputs a $d=256$ dimension vector representation for each of the $M$ input tracks.
This is followed by a sinusoidal positional encoding \cite{VaswaniAttention} of track impact parameters and angular displacements from the jet axis.
The positional encoding allows the self-attention mechanism to readily identify nearby tracks without the computational overhead of adding explicit edge features as suggested in \rcite{ParT}, which improves performance.
The input embeddings are then fed into a 4-layer self-attention Transformer encoder network \cite{VaswaniAttention}, which performs feature extraction and outputs a second embedding for each track.
This Transformer encoder has a model dimension $d$ and a feed-forward update width of $2d$.
Rectified Linear Unit (ReLU) activations are used throughout the model.

\subsection{Object Decoder}
The object decoder computes an explicit representation of a set of \textit{object queries}, which represent possible output vertices.
At the start of the decoding stage, object queries are initialised as a set of $N=5$ learned vectors of dimension~$d$.
The number of object queries is chosen as the maximum number of target vertices per jet in the training sample, but could easily be adapted for other applications.
In a given decoder layer, each object query is updated via information exchange with the updated track representations from the encoder (via cross-attention) and other object queries (via self-attention), enabling the model to efficiently aggregate information about the entire jet.
For the cross-attention operations, we use the MaskAttention operator~\rcite{mask2former} to sparsify the attention matrix based on intermediate mask predictions.
This operator enables object queries to attend only to relevant tracks and improves performance.
During the decoding stage, the track representations are also updated by attending to the object queries.
We use a three layer decoder.
The output of the final layer is the set of updated track representations and the set of updated object queries, which are used to predict the presence and properties of any vertices, along with their track masks, as described in the following.

\subsection{Task Heads}\label{sec:tasks}
A series of task-specific dense networks are used to produce the various model outputs.
Each network has three hidden layers composed of 256, 256, and 128 nodes respectively.
More details about each type of output are provided below.

\subsubsection{Vertex Class}
Vertex class labels are either \bhadron, \chadron, or \NULL, with the latter class indicating that no target vertex is present.
If the model predicts an object query as \NULL, other outputs for that object are ignored.
The other valid (i.e. not \NULL) object queries are then interpreted as output reconstructed vertices, enabling the model to reconstruct a variable number of vertices.
For example, since we consider only jets instantiated from a single quark, \chadrons inside \bjets can be separately reconstructed and exploited in downstream efforts.

\subsubsection{Masks}
The assignment of input tracks to output vertices is provided by a series of $N$ one-dimensional mask arrays, each of which is of length $M$.
In order to obtain the masks, a single layer perceptron is first used to compute a mask token for each object query which has the same dimension, $d$, as the track representation.
The masks for each vertex are then computed as binarised dot products between the mask token and the output track embeddings produced by the decoder.
A value of 1 in the $i$th position of mask $j$ indicates that the $i$th input track is assigned to vertex $j$, while a 0 indicates no assignment.

\subsubsection{Vertex Properties}
Vertex fitting is implemented as a dense, multi-target regression network with three output nodes; one for each regression target.
We use simulation-level properties of the heavy-flavour hadrons to define these targets.
While there is flexibility to add predictions for a wide range of properties, three have been chosen for this study: the hadron's transverse momentum (\pt), the radial flight length of the hadron with respect to the beamline (\lxy), and the angular displacement between the jet axis and the hadron (\dr).
Since each regression target is always positive, a ReLU activation is applied to the output layer.
The same network is used to predict properties for each object query.

\subsubsection{Jet Flavour Prediction}
The model outputs a probability associated with each possible jet class: $b$-jet, $c$-jet, or light-jet.
In order to perform jet flavour classification, the track and vertex representations are pooled independently using attention pooling. 
The resulting pooled representations are concatenated and fed into a fourth and final dense network task which outputs the jet flavour probabilities.

\subsection{Losses}
The loss function $L$ is made up of an aggregate loss term for the vertexing tasks $L_{\text{Vertex}}$ and a categorical cross entropy (CE) loss for the jet flavour prediction $L_{\text{Jet Flavour}}$ as shown in the equation below:
{
\begin{equation}
    L =
    L_{\text{Vertex}} +
    L_{\text{Jet Flavour}}.
\end{equation}
}

The vertexing loss is composed of several terms.
The first of these is a categorical CE loss for the vertex class labels $L_{\text{Class}}$, including the \NULL label.
For the masks, a linear combination of a binary CE loss $L_{\text{BCE}}$ and a dice loss~\cite{2017arXiv170703237S} $L_{\text{Dice}}$ are applied.
Finally, for the regression, a mean absolute error loss ${L_{\text{Regression}}}$ is applied for each of the vertex properties being reconstructed.
The vertexing loss function is therefore as follows:
{
\begin{equation}
\begin{split}
    L_{\text{Vertex}} = & \; 0.5 L_{\text{Class}} +
    2 L_{\text{Regression}} \\
    & + \underbrace{12 L_{\text{BCE}} + 8 L_{\text{Dice}}}_{L_{\text{Mask}}},
\end{split}
\end{equation}
}
where weights have been applied to each term to produce an acceptable trade-off between the performance of the different tasks.
The optimal bipartite matching between predictions and targets is used to define the vertexing loss following the procedure described in \rcite{2020arXiv200512872C}.


\section{Dataset \& Experimental Setup}
\label{sec:experiments}

\subsection{Dataset}\label{sec:dataset}

Samples of hadronically decaying $t\overline{t}$ events from simulated $pp$ collisions are generated using \textsc{Pythia 8.307}~\cite{bierlich2022comprehensive} at a centre-of-mass energy of $\sqrt{s} = 13$ TeV with an average of $50$ interactions per bunch crossing.
The detector response is simulated with \textsc{Delphes 3.5.0}~\cite{de_Favereau_2014} using a modified version of the ``ATLAS Pileup'' configuration.
The modifications are intended to make the simulation more realistic, and are listed below.
Firstly, the track reconstruction efficiency parameterisation was modified to more closely match current ATLAS tracking software performance as measured in \rcite{ATLAS:2023iat}.
Secondly, parameterised impact parameter smearing is applied based on impact parameter resolutions obtained by ATLAS after the addition of the Insertable B-Layer in 2015~\cite{ATLAS-TDR-19}.
Finally, smearing is applied to track $\eta$ and $\phi$, based on the ATLAS Inner Detector Technical Design Report~\cite{ATLAS-TDR-04}.
Even with these modifications, several limitations remain, for example the interaction of displaced charged hadrons with the detector material is not simulated.

Jets are reconstructed using \textsc{FastJet} \cite{fastjet} with a radius parameter of $R=0.4$.
Flavour labels are assigned based on the presence of heavy-flavour hadrons with $\pt>5$~GeV within $dR < 0.4$ of the jet axis:
\bjets contain \bhadrons, \cjets contain \chadrons but no \bhadrons, and \ljets contain neither.
Training jets must satisfy with $20 < \pt < 250$~GeV and $|\eta| < 2.5$ and are selected to balance \pt and $\eta$ distributions across flavours.
Heavy-flavour with $\pt > 5$~GeV are used as targets for vertex reconstruction.

The full dataset used for training and evaluation is available in Ref. \cite{delphes_dataset}.

\subsection{Inputs \& Training Setup}\label{sec:train-setup}

The model takes as inputs the 50 highest \pt tracks associated to each jet.
This number is chosen such that the all the tracks are included from vast majority of jets.
For each track, the following variables are used as inputs: the signed impact parameters ($d_0$ and $z_0$), the fraction of the jet \pt which is carried by the track, and track $\eta$ and  $\phi$ relative to the jet axis.
The signed transverse impact parameter $d_0$ is the shortest distance in the transverse plane from the primary vertex to a track.
The sign of $d_0$ indicates whether the track is on the same side as or the opposite side of the primary vertex relative to the particle's momentum direction.
The longitudinal impact parameter $z_0$ is the distance along the beam axis (z-axis) from the primary vertex to the point where the track's transverse impact parameter is measured.
The jet \pt and $\eta$ are also concatenated to each track.
A total of $13.5$~million jets are used for training, equally split between the three jet flavours.
Two further sets of $1.35$~million jets are used for each of validation and testing.

Models are trained on 3 NVIDIA A100 GPUs using a combined batch size of 6000 jets.
Full training of the MaskFormer model was completed in 11 hours.
The AdamW \cite{loshchilov2019decoupled} optimiser is used with a weight decay of $1\times10^{-5}$ and a OneCycle learning rate schedule \cite{smith2018superconvergence} with an initial warm-up learning rate of $1\times10^{-7}$, peaking at $5\times10^{-4}$ once $5\%$ of training steps have passed, before decreasing to a final value of $1\times10^{-5}$. 
Models are trained for 50 epochs and evaluated at the epoch with the lowest validation loss.


\section{Results}\label{sec:results}

We evaluate MaskFormer's performance in three key areas: vertex finding, vertex fitting, and flavour tagging.
For a vertex finding and flavour tagging benchmark, we trained an EdgeClassifier model as described in \cref{sec:related-hep}.
The EdgeClassifier model is similar to the GN1 tagger developed at ATLAS \cite{GN1}, which has been shown to have superior vertex finding performance to procedural vertexing algorithms \cite{svsthesis}.
The benchmark EdgeClassifier model is trained on the same input jets and has a similar architecture to the \maskformer model (without the object decoder).
The embedding dimension of the Transformer encoder for the EdgeClassifier is doubled to $d'=512$ compared to the \maskformer such that both models have 6.3 million trainable parameters.
The inference times of the two models, as measured on a CPU using ONNX \cite{bai2019}, are comparable, each taking approximately 5~ms to process a single jet.

\subsection{Vertex Finding}

\begin{figure*}[!ht]
    \centering
    \begin{subfigure}[b]{0.45\textwidth}
        \centering
        \includegraphics[width=\textwidth]{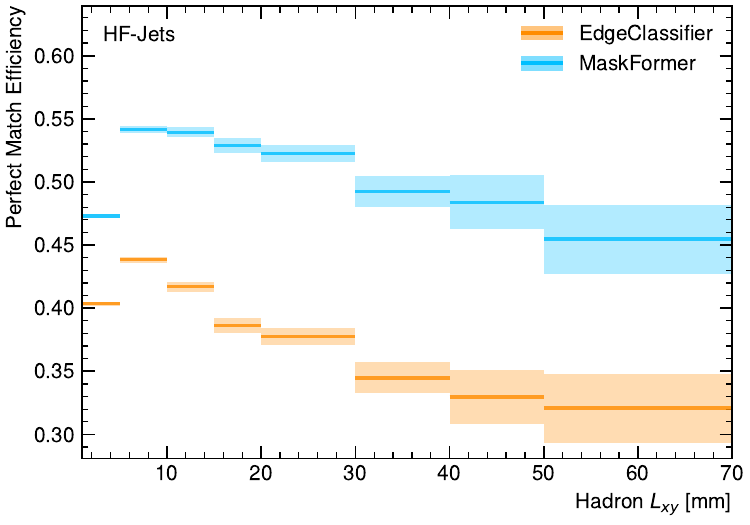}
    \end{subfigure}%
    \hspace{2em}%
    \begin{subfigure}[b]{0.45\textwidth}
        \centering
        \includegraphics[width=\textwidth]{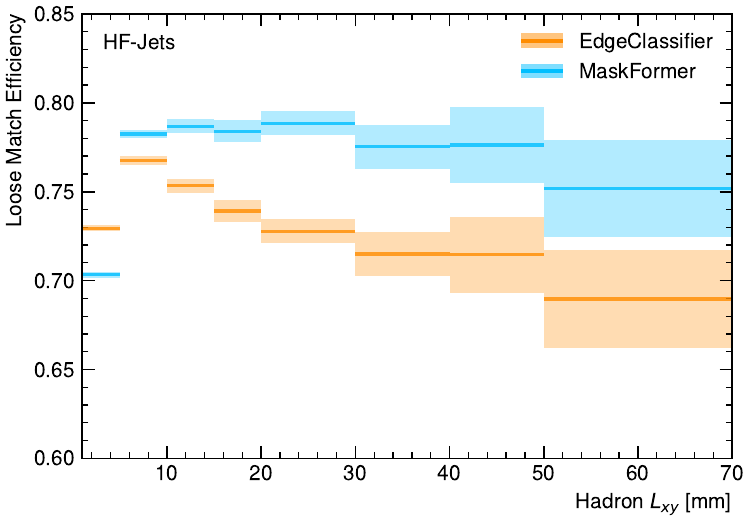}
    \end{subfigure}%
    \caption{
        Heavy-flavour hadron vertex reconstruction efficiency as a function of hadron \lxy for vertices in \bcjets (HF-jets) for the perfect match (left) and loose match (right). 
        Only vertices with at least two tracks at truth level are considered.
        The shaded region represents binomial error bands.
    }
    \label{fig:vert-eff}
\end{figure*}

Vertex finding is the task of correctly assigning tracks to vertices.
In order to measure successful vertex finding, we define two matching criteria; \textit{perfect} and \textit{loose}, which are based on the recall and purity of tracks assigned to a given predicted vertex.
The perfect matching requires predicted vertices to have a recall and purity of 100\%, while the loose criteria requires a minimum recall of 50\% and minimum purity of 50\%, where recall and purity are defined below.
{
\begin{align}
    \text{Recall} &= \frac{N_{\text{Correctly Associated Tracks}}}{N_{\text{Tracks in Target Vertex}}}, \\
    \text{Purity} &= \frac{N_{\text{Correctly Associated Tracks}}}{N_{\text{All Associated Tracks}}}.
\end{align}
}
The EdgeClassifier reconstructs vertices from compatible track pairs, limiting it to only identifying vertices with two or more associated tracks as training targets.
While the \maskformer has no such intrinsic limitation, we consider only vertices with at least two tracks associated at truth level to facilitate comparison.
In addition, hadrons with $\lxy < 1$ mm at simulation level are excluded from the evaluation as they are not sufficiently displaced from the primary vertex to avoid additional reconstruction effects.

The heavy-flavour vertex reconstruction efficiency as a function of hadron \lxy is shown in \cref{fig:vert-eff} for both the perfect and loose matching criteria.
The results include all vertices within heavy-flavour jets: \bhadron and $b\rightarrow$ \chadron vertices (inside \bjets) and \chadron vertices (inside \cjets).
For the perfect matching criteria, \maskformer shows an absolute efficiency increase of approximately \pct{10} compared to the EdgeClassifier over the full \lxy range.
This corresponds to a relative improvement of \pct{20}-\pct{40}.
In addition, \maskformer is able to find more than \pct{75} of vertices with $\lxy > 5$ mm with the loose matching criteria, and displays a reduced dependence on \lxy.
It should be noted that at higher \lxy values (above approximately 30~mm), heavy-flavour hadrons will decay in the instrumented detector volume in most general-purpose detectors and have considerable momentum. This typically leads to various effects that negatively impact track reconstruction, such as the absence of hits or incorrect hits in the innermost layers and a failure to reconstruct close-by tracks. Since these effects are not fully modelled in \textsc{Delphes}, further studies using a high-fidelity simulation will be necessary to accurately assess performance in this region.

Since the model is trained to reconstruct only heavy-flavour vertices, the separation of heavy-flavour vertices and other displaced vertices (for example from $s$-hadrons or conversions) is implicitly learned.
The fake rate  (i.e. the rate at which the model predicts a heavy-flavour vertex when none exists) therefore includes a contribution from combinatorial fakes and also from the misidentification of non-heavy-flavour vertices.
To assess the fake rate, we analyse the heavy-flavour vertex rate in \ljets.
Since \ljets contain no heavy-flavour hadrons by definition, any reconstructed vertices can be considered as false positives.
We observe a fake rate of approximately \pct{5} that is consistent for both models within the statistical uncertainty, and which remains broadly constant as a function of jet \pt.

A single jet can contain multiple vertices due to the presence of tertiary decays from heavy-flavour hadrons such as a \bhadron decaying to a \chadron.
Although \maskformer is able to maintain a high vertex finding efficiency in these cases, a reduction is observed as the heavy-flavour hadron multiplicity increases.
Vertices in jets with containing single hadron are reconstructed with a loose match efficiency of \pct{79}, which drops slightly to \pct{77} for vertices within jets with exactly two hadrons (typically \bjets with one \chadron).
Meanwhile, vertices in jets with $\geq 3$ hadrons (typically \bjets with two \chadrons) are reconstructed with \pct{60} efficiency.
Notably, MaskFormer performs similarly for both \bhadron and \chadron vertices, demonstrating its robustness across different types of heavy-flavour decays.

\maskformer can assign the correct track to a hadronic vertex with only one single reconstructed decay product with an efficiency of 48\%.
The rate of single track vertices in \ljets is 0.055.
This can be compared with a loose match efficiency (rate in \ljets) of 74\% (0.048) for vertices with two or more associated tracks at truth level.

\begin{figure*}[!h]
    \centering
    \begin{subfigure}[b]{0.33\textwidth}
        \centering
        \includegraphics[width=\textwidth]{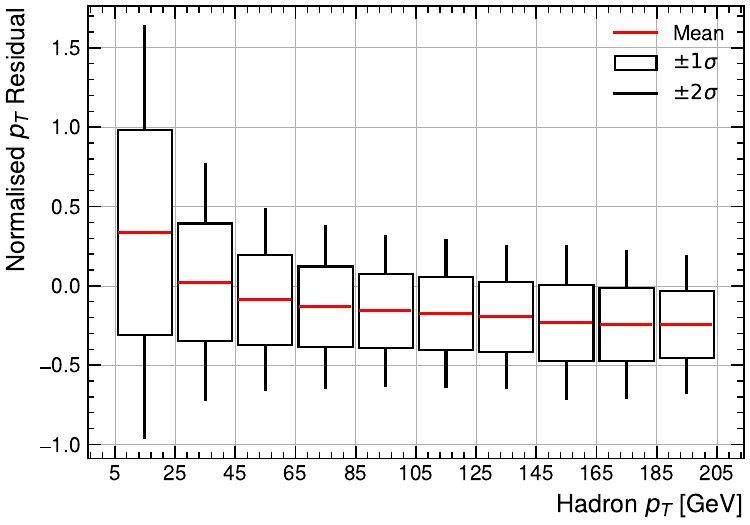}
        \label{fig:vert_fit_pt}
    \end{subfigure}%
    \begin{subfigure}[b]{0.33\textwidth}
        \centering
        \includegraphics[width=\textwidth]{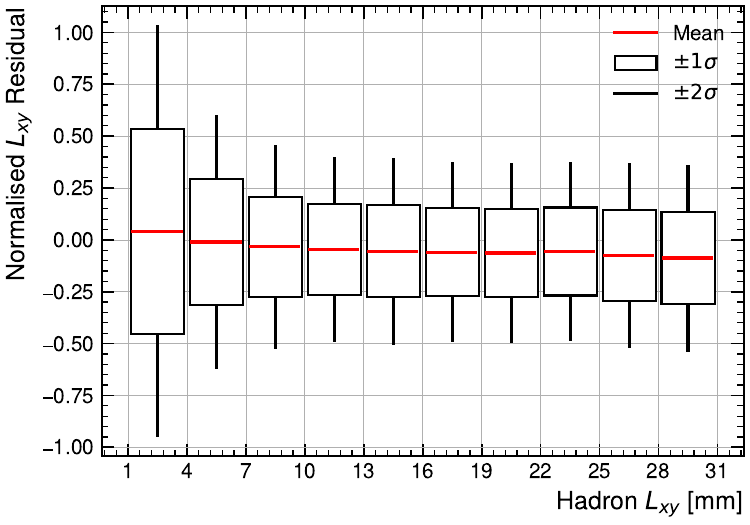}
        \label{fig:vert_fit_lxy}
    \end{subfigure}%
    \begin{subfigure}[b]{0.33\textwidth}
        \centering
        \includegraphics[width=\textwidth]{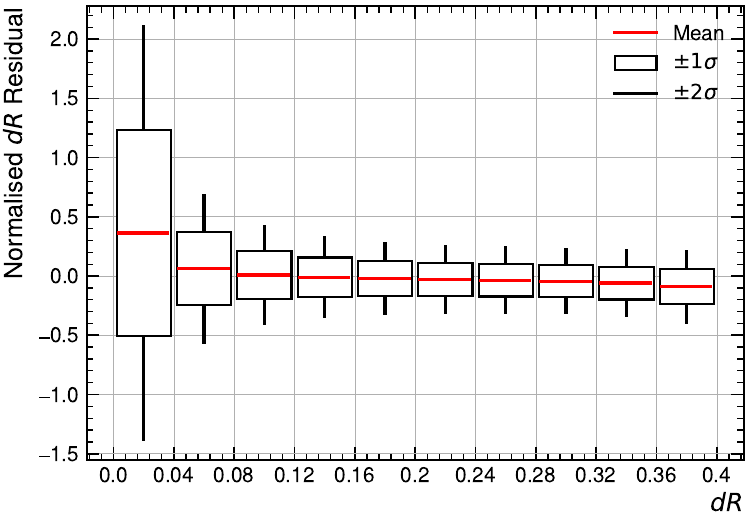}
        \label{fig:vert_fit_dr}
    \end{subfigure}%
    \caption{
    Normalised fit residuals $( x_{\text{Pred}} - x_{\text{True}} ) / x_{\text{True}}$ for \pt (left), \lxy~(middle), and \dr~(right) as a function of the respective target properties for all reconstructed hadrons.
    The mean-normalised residual is shown by the red line, while the box and whiskers demarcate the region containing $\pct{68}$ and $\pct{95}$ of the data respectively.
    }
    \label{fig:vert-fit}
\end{figure*}

\begin{figure*}[!h]
    \centering
    \begin{subfigure}[b]{0.33\textwidth} 
        \centering
        \includegraphics[width=\textwidth]{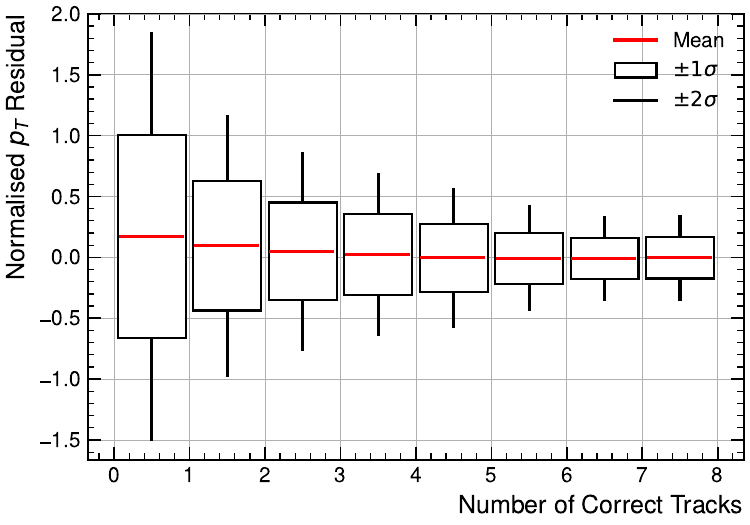}
    \end{subfigure}%
    \begin{subfigure}[b]{0.33\textwidth} 
        \centering
        \includegraphics[width=\textwidth]{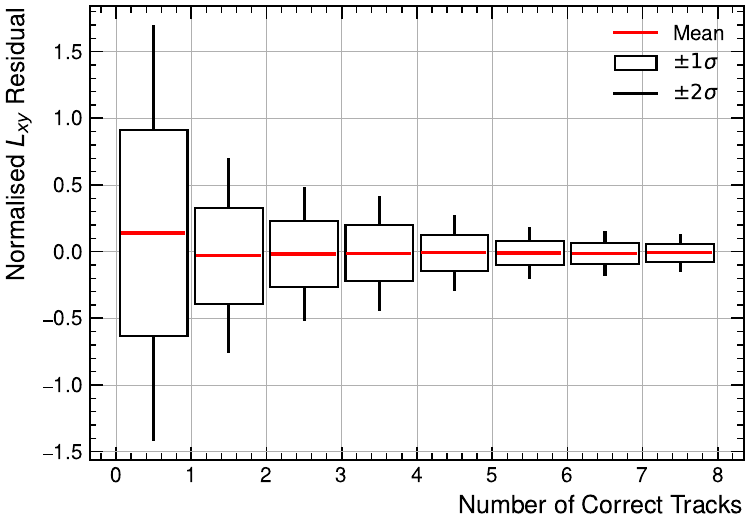}
    \end{subfigure}%
    \begin{subfigure}[b]{0.33\textwidth} 
        \centering
        \includegraphics[width=\textwidth]{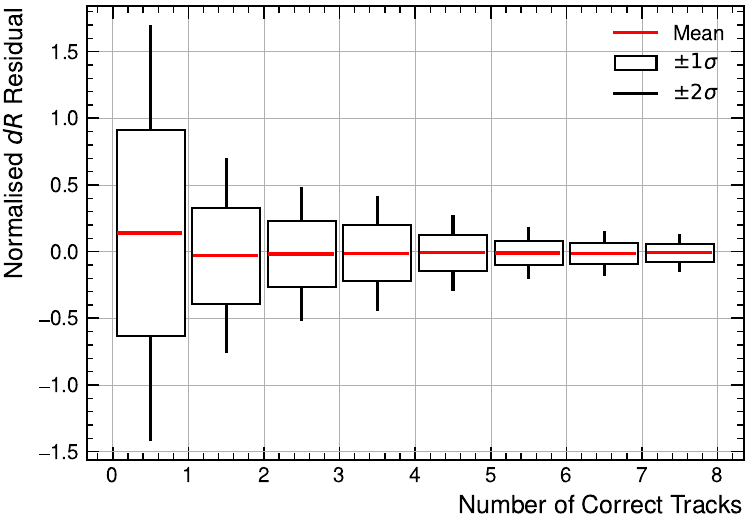}
    \end{subfigure}%
    \caption{
    Normalised fit residuals $( x_{\text{Pred}} - x_{\text{True}} ) / x_{\text{True}}$ for \pt (left), \lxy (middle), and $\Delta R(\text{Jet, Hadron})$ (right) as a function of the number of correctly associated tracks to the hadronic vertex for all reconstructed hadrons.
    The mean-normalised residual is shown by the red line, while the box and whiskers demarcate the region containing $\pct{68}$ and $\pct{95}$ of the data respectively.
    }
    \label{fig:vert-eff-per-track}
\end{figure*}

\subsection{Vertex Fitting}

Vertex fitting is the estimation of the properties of a vertex from its constituent tracks.
As described in \cref{sec:tasks}, MaskFormer simultaneously fits the properties of all vertices in the jet via a series of regressions to simulation-level quantities of heavy-flavour hadrons.
As discussed, this is a key improvement over the EdgeClassifier, which is unable to produce an in-model estimate of the properties of reconstructed vertices.

The normalised residuals of the MaskFormer vertex fit are shown in \cref{fig:vert-fit} for vertex \pt, \lxy, and \dr, as a function of the respective target.
The results are shown for all reconstructed vertices without applying quality requirements on the assigned tracks.
For consistency, hadrons with $\lxy < 1$ mm at simulation level are excluded from the evaluation, although this has a negligible impact on the regression performance for \pt and \dr.

Normalised residuals are largest at low values of the target distribution, but are otherwise generally consistent in the bulk of the distributions.
For each target, and especially for \pt, the model tends to overestimate below the mean and vice versa.
Further work is required to reduce this bias towards predicting the central values of each distribution.

Normalised residuals are also shown for \pt, \lxy, and \dr as a function of the number of correctly assigned tracks in the predicted vertex in \cref{fig:vert-eff-per-track}.
For all variables, the accuracy of the prediction has a significant positive correlation with the number of correctly assigned tracks, and the residuals are largest when no tracks are correctly assigned.
The number of correctly assigned tracks required for \pct{95} of vertices to have predictions within \pct{50} of their target value is five for the prediction of \pt, but only two for the prediction of \lxy, reflecting the fact that it is easier to recover the hadron \lxy with fewer tracks.

Overall, \maskformer displays a promising ability to accurately recover the properties of hadronic vertices inside jets.
We also find that the regressed outputs from the model for \dr and \pt are more accurate than the estimate obtained from a simple 4-vector sum of the tracks assigned to the vertex.
We hypothesise that a learned vertex fit is advantageous as it can incorporate global information about the jet and account for incorrectly assigned, mis-reconstructed, and un-reconstructed tracks.
Maskformer is also able to account for the neutral component of the heavy-flavour hadron decay, which is not possible with a vertex fitting approach based on a fit of charged particle tracks only.
As a result, our approach is able to reconstruct the simulation-level \pt of the heavy-flavour hadron with much greater accuracy than using a 4-vector sum.

\subsection{Flavour tagging}

Previous approaches to flavour tagging have shown that the inclusion of information about secondary vertices improves heavy-flavour jet identification performance \cite{ATLAS:2019bwq,CMS-BTV}.
As discussed in \cref{sec:intro}, recent flavour tagging models from ATLAS \cite{GN1} have included an ML-based vertex reconstruction component which is able to find (but not fit) secondary vertices.
This approach is similar to the EdgeClassifier benchmark used in this study.
In this work, we complete the transition to fully ML-based vertex reconstruction which is capable of finding and fitting an arbitrary number of secondary vertices inside a jet and is combined with a prediction of the jet flavour.

\begin{figure}[!htb]
    \centering
    \includegraphics[width=0.45\textwidth]{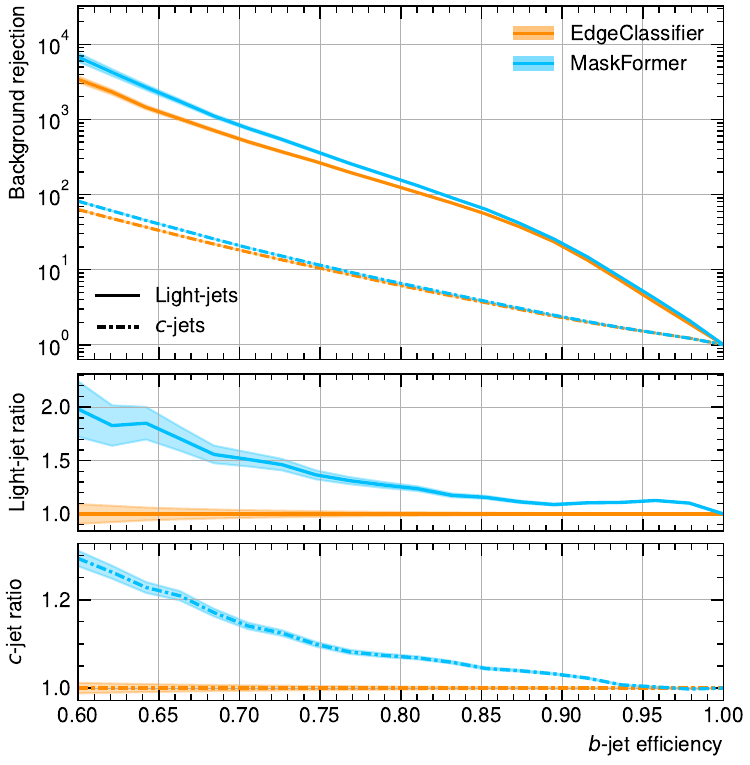}
    \caption{
    Background rejection as a function of \bjet selection efficiency for \ljets (solid) and \cjets (dashed).
    Binomial errors are shown as shaded bands.
    }
    \label{fig:ftag}
\end{figure}

In \cref{fig:ftag}, the flavour-identification performance of the EdgeClassifier baseline and \maskformer models is shown.
Both models classify jets using a discriminant following the procedure used in \rcite{GN1}.
A significant improvement in background rejection is observed for the \maskformer,
with \pct{50} improvement in \lrej and \pct{15} improvement in \crej for a \bjet selection efficiency of \pct{70}.
This result demonstrates that the \maskformer architecture, which is able to fully reconstruct the heavy-flavour vertices inside jets, is better able to extract information relevant to the jet flavour from the available training data.
Future studies will include the use of more realistic detector simulations and increased training statistics.

\section{Conclusion}\label{sec:conclusion}

We present a novel approach for reconstructing secondary vertices inside jets based on a MaskFormer neural network architecture.
This approach enables the simultaneous finding and fitting of an arbitrary number of secondary vertices within jets.
Our results demonstrate a significant improvement in vertex finding efficiency over an existing state-of-the-art vertexing algorithm while maintaining a comparable fake rate.
Crucially, \maskformer also enables the prediction of vertex properties via a class label prediction and a set of regression tasks.
This more streamlined and versatile approach to vertex reconstruction is readily adaptable to new environments and datasets.

When flavour tagging, MaskFormer achieves a notable enhancement in performance, with a 50\% improvement in light-jet rejection and a 15\% improvement in \textit{c}-jet rejection at a \textit{b}-jet selection efficiency of 70\%.
This demonstrates the model's improved capability to extract relevant information for jet classification, paving the way for more accurate and efficient object identification in high-energy physics experiments.

Looking ahead, we plan to extend the deployment of this vertex reconstruction tool to more complex environments, such as boosted large-radius jets, and across full events.
The capability to fully reconstruct all the heavy-flavour hadrons in an event could lead to the development of Infrared and Collinear (IRC) safe flavour tagging algorithms, for example following the approach in Ref. \cite{Caola:2023wpj}, however further studies would be required to first quantify the degree of IRC safety in the vertex reconstruction step.
While this work focused on the reconstruction of heavy-flavour decay vertices, exploring the reconstruction of primary $pp$ interaction vertices and secondary vertices from light-flavour hadrons is an important next step.

Due to it's flexibility, MaskFormer also has the potential to be repurposed for other many-to-many reconstruction tasks in particle physics, such as track reconstruction or the reconstruction of complex events with multiple heavy resonances.

In conclusion, our work not only enhances the current capabilities in vertex reconstruction and flavour tagging but also lays the groundwork for future innovations in the field, offering a powerful and flexible tool for reconstruction high-energy physics experiments.

\clearpage
\backmatter




\bibliography{references}

\end{document}